\documentclass[12pt,twoside]{article}
\usepackage{amsmath,amssymb,amsfonts}
\usepackage{graphicx}
\usepackage{hyperref}
\hypersetup{hidelinks=true}
\usepackage{amsmath,amssymb,amsfonts}
\usepackage{graphicx}
\usepackage{textcomp}

\usepackage{caption}
\usepackage{subcaption}
\usepackage[linesnumbered, ruled]{algorithm2e}
\usepackage{booktabs}
\usepackage{amsmath}
\usepackage{amssymb}
\usepackage{multirow}
\usepackage{makecell}
\usepackage{hyperref}
\usepackage{textcomp}

\newcommand{\keywords}[1]{\textit{Keywords:} #1}

\usepackage{hyperref}

\usepackage{fancyhdr}		

\renewcommand{\thesection}{{\sf \Roman{section}}.}
\renewcommand{\thesubsection}{\thesection{\sf \Alph{subsection}}.}
\renewcommand{\thesubsubsection}{\thesubsection{\sf \arabic{subsubsection}}.}

\newcommand{\cen}[1]{\begin{center} #1 \end{center}}

\usepackage[mathlines]{lineno}

\usepackage{hyperref}
\hypersetup{ colorlinks,
    citecolor=blue,
    filecolor=blue,
    linkcolor=blue,
    urlcolor=blue
}

\usepackage{xcolor}
\definecolor{gray}{rgb}{0.6,0.6,0.6}
\definecolor{red}{rgb}{0.85,0,0}
\definecolor{green}{rgb}{0,0.85,0}
\definecolor{blue}{rgb}{0,0,0.85}
\definecolor{beige}{rgb}{0.92,0.87,0.78}
\usepackage[all]{hypcap}    

\begin{document}

\cen{\sf {\Large {\bfseries FDDM: Unsupervised Medical Image Translation with a Frequency-Decoupled Diffusion Model} \\  
\vspace*{10mm}
Yunxiang Li\textsuperscript{a}, Hua-Chieh Shao\textsuperscript{a}, Xiaoxue Qian\textsuperscript{a}, You Zhang\textsuperscript{a*}
} \\

\textsuperscript{a}Department of Radiation Oncology, UT Southwestern Medical Center, Dallas, 75390, TX, USA \\

\textsuperscript{*}Author for correspondence: you.zhang@utsouthwestern.edu \\

}

\setcounter{page}{1}
\pagestyle{plain}

\begin{abstract}
Diffusion models have demonstrated significant potential in producing high-quality images in medical image translation to aid disease diagnosis, localization, and treatment. Nevertheless, current diffusion models have limited success in achieving faithful image translations that can accurately preserve the anatomical structures of medical images, especially for unpaired datasets. The preservation of structural and anatomical details is essential to reliable medical diagnosis and treatment planning, as structural mismatches can lead to disease misidentification and treatment errors. In this study, we introduce the Frequency Decoupled Diffusion Model (FDDM) for MR-to-CT conversion. FDDM first obtains the anatomical information of the CT image from the MR image through an initial conversion module. This anatomical information then guides a subsequent diffusion model to generate high-quality CT images. Our diffusion model uses a dual-path reverse diffusion process for low-frequency and high-frequency information, achieving a better balance between image quality and anatomical accuracy. We extensively evaluated FDDM using public datasets for brain MR-to-CT and pelvis MR-to-CT translations, demonstrating its superior performance to other GAN-based, VAE-based, and diffusion-based models. The evaluation metrics included Fréchet Inception Distance (FID), mean absolute error (MAE), mean squared error (MSE), and Structural Similarity Index Measure (SSIM). FDDM achieved the best scores on all metrics for both datasets, particularly excelling in FID, with scores of 25.9 for brain data and 29.2 for pelvis data, significantly outperforming other methods. These results demonstrate that FDDM can generate high-quality target domain images while maintaining the accuracy of translated anatomical structures.
\end{abstract}

\begin{keywords}
Medical Image Translation, Diffusion Model, Generative Model
\end{keywords}

\newpage

\section{Introduction}
Magnetic Resonance (MR) imaging and Computed Tomography (CT) imaging play pivotal and complementary roles in medical diagnosis and treatment planning \cite{kazemifar2019mri,payne2006applications,thornton1992clinical,goitein1979value,saw2005determination,shao2023real,zhang2023children,shao2023real2}. MR offers exceptional soft-tissue contrast and CT is usually preferred for lung and bony anatomy scans. For radiation therapy, MR scans are often used for tumor localization and segmentation, while CT scans provide electron density information for calculating radiotherapy plan doses \cite{chandarana2018emerging,edmund2017review,wang2022recurrence}. Current clinical practices usually acquire MR and CT as separate scans to serve different needs, incurring additional imaging time, imaging radiation dose (from CT), and medical costs. These sequential acquisitions also introduce MR/CT misregistrations that affect the alignment and localization of anatomies.  The prospect of acquiring a single modality (for instance, MR), and converting it to another modality (CT) via numerical methods (MR-to-CT translation) can fundamentally address the above-mentioned challenges. However, such MR-to-CT translation remains challenging due to substantial differences between MR and CT scans in terms of image intensities and features.

\begin{figure*}[ht]
\centering
  \includegraphics[width=.99\textwidth]{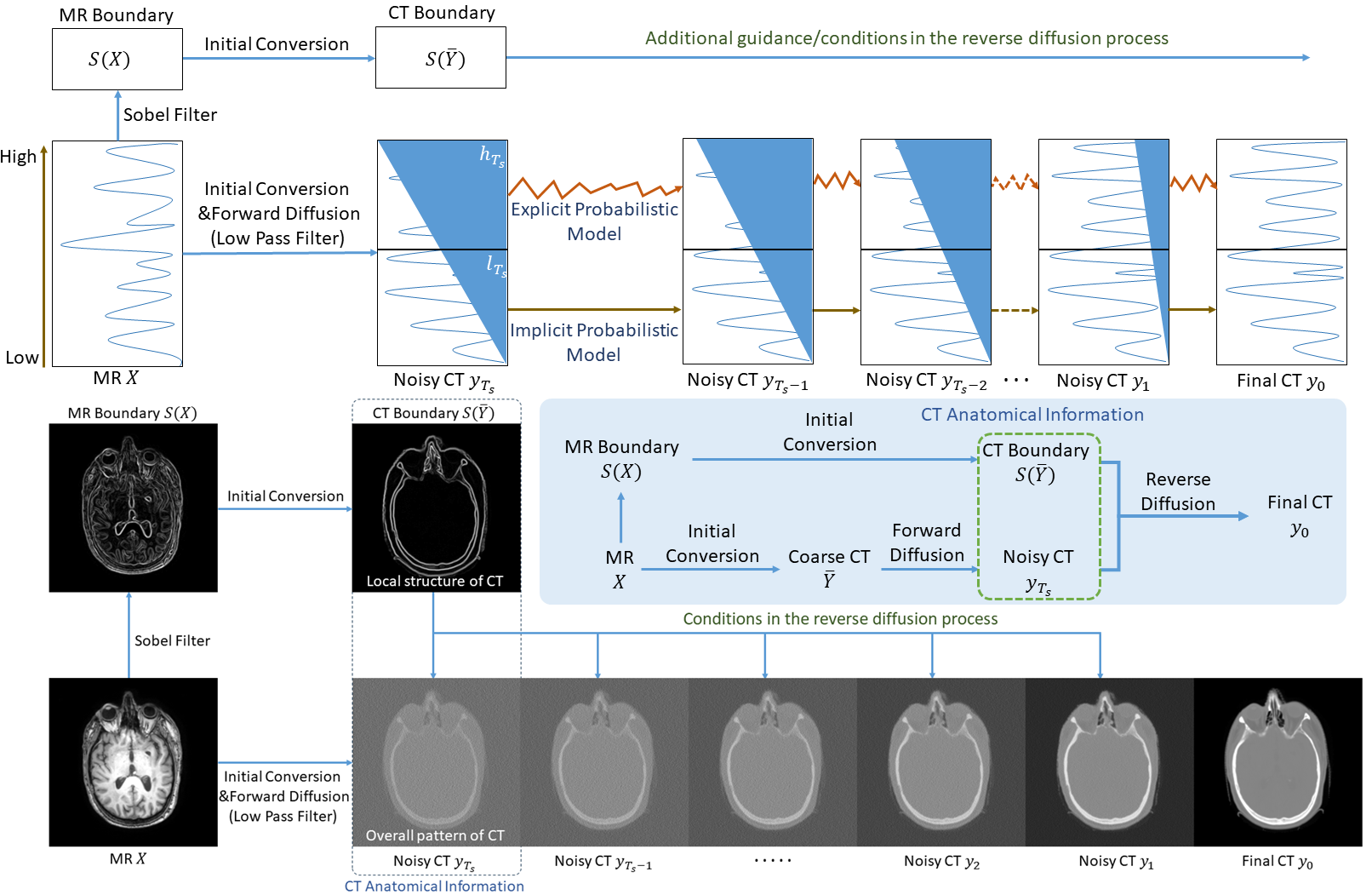}
  \caption{Overview of the FDDM framework for MR-to-CT conversion. We first extract the boundary $S(X)$ of the MR image $X$, and then input the MR image $X$ along with its boundary $S(X)$ into the initial conversion module to obtain a coarse CT $\overline{Y}$ and CT boundary $S(\overline{Y})$. Subsequently, the overall pattern information of CT, $y_{T_s}$, is obtained by subjecting the coarse CT image $\overline{Y}$ to a forward diffusion process (low-pass filter). The CT boundary $S(\overline{Y})$ and noisy CT $y_{T_s}$(overall pattern information) serve as anatomical information to jointly guide the reverse diffusion process to generate the final CT $y_0$.}
  \label{fig:brief}
\end{figure*}

The current mainstream image translation methods are mostly based on generative adversarial networks (GAN)~\cite{goodfellow2020generative} or variational autoencoders (VAE)~\cite{kingma2019introduction}. These methods show advantages in preserving the anatomical structures (high faithfulness). However, they are subject to many issues including pattern collapse, premature discriminator convergence, or resolution degradation,  affecting the quality of the translated images (low realism). Recently, diffusion models demonstrated superior generative capabilities, achieving impressive Frechet Inception Distance scores (FID) that measure the realism of generated images~\cite{heusel2017gans,dhariwal2021diffusion,yang2022diffusion,song2021scorebased,song2021denoising}. A representative diffusion framework is the Denoising Diffusion Probabilistic Model (DDPM)~\cite{ho2020denoising}, which utilizes a Markov chain Monte Carlo (MCMC) process to progressively add Gaussian noises to images (forward diffusion),  followed by a learned reverse diffusion process that maps noises back to images. However, despite their potential, diffusion models have found significant challenges in medical image applications, particularly in preserving and maintaining the integrity of anatomical structures during image translations (low faithfulness). The forward diffusion process, where random Gaussian noises are continuously added, gradually obliterates structural details of high spatial frequencies. Such structural information is crucial for medical diagnosis/treatment; and is difficult for the reverse diffusion process to completely restore. In general, diffusion models, while excelling in creating highly realistic images, often struggle to maintain the faithfulness of translated anatomical structures. It is especially difficult to develop diffusion models on unpaired imaging datasets, which lack image pairs of well-corresponded anatomy to serve as additional clues for network training. However, unpaired datasets are far more prevalent and easier to obtain in medicine than paired datasets. Overall, models based on VAE or GAN can achieve unsupervised image translation, but the image quality is often subpar. Diffusion models can synthesize high-quality images but struggle with unsupervised training. Therefore, there is a pressing need to develop a diffusion model that can accurately perform image-to-image translation from unpaired datasets (unsupervised learning). In this work, we propose a novel framework named the Frequency Decoupled Diffusion Model (FDDM) for MR-to-CT conversion. FDDM is a two-stage framework: the first stage uses a VAE to achieve unsupervised MR-to-CT anatomical information translation, and the second stage uses a diffusion model to generate high-quality CT images conditioned on the anatomical information. The advantage of our approach is that it combines the unsupervised learning and anatomical structure preservation capability of the VAE with the high-quality image synthesis capability of the diffusion model, allowing each stage of the model to focus on its strengths. Additionally, we separate the low-frequency and high-frequency components during the reverse diffusion process, generating them through different models to achieve a balance between image quality and anatomical accuracy.

In summary, our key contributions are: \\
\textbf{1.} We introduce a novel Frequency Decoupled Diffusion Model (FDDM) for unsupervised MR-to-CT image translation. FDDM achieves high anatomical accuracy (faithfulness) and high image quality (realism) through a two-stage design that first translates anatomical information and then uses the diffusion model to conform to the CT data distribution.\\
\textbf{2.} To achieve high-accuracy conversions, we optimized both the forward and reverse diffusion processes. The forward diffusion uses blue noise to approximate low-pass filtering, obtaining the low-frequency information of the CT as the starting point for reverse diffusion. During the reverse diffusion process, we generate high-frequency and low-frequency components through tailored models, to promote high fidelity for low frequencies and good alignment with the CT data distribution for high frequencies. \\
\textbf{3.} Compared with existing state-of-the-art methods (including GAN-based, VAE-based, and other diffusion model-based methods), FDDM demonstrates superior performance in MR-to-CT translation on both brain and pelvis datasets.

\section{Related Work}
\subsection{Unsupervised Medical Image Translation.}
Image-to-image translation has been widely employed in medical applications, including but not limited to cross-modality image registration, low-dose CT denoising, fast MR reconstruction, and metal artifact reduction \cite{armanious2020medgan,kaji2019overview}. For medical image translation, it is critical to preserve the integrity of anatomical structures when converting images between domains. The primary methods including Generative Adversarial Networks (GANs) ~\cite{goodfellow2020generative} and Variational Autoencoders (VAEs) generally perform well in structure retention (high faithfulness)~\cite{kingma2019introduction}. Notable developments in GAN-based methods include CycleGAN~\cite{zhu2017unpaired} which promotes cyclic consistency; and GcGAN~\cite{fu2019geometry} for unilateral, geometry-consistent mapping. Additionally, RegGAN~\cite{kong2021breaking} blends image translation with registration. Despite the demonstrated success, GANs often face training instabilities that reduce the quality of generated images (low realism). In addition to GANs, VAE-based approaches like UNIT~\cite{liu2017unsupervised} and MUNIT~\cite{huang2018multimodal} share similar advantages/disadvantages.

\subsection{Diffusion Models for Image Translation}
Recently, diffusion models have emerged as powerful tools for generating high-quality and realistic images, surpassing GANs and VAEs. One example of the diffusion model is SDEdit~\cite{meng2021sdedit}, which translates images through iterative denoising via stochastic differential equations. Another diffusion model, SynDiff~\cite{ozbey2023unsupervised} utilizes CycleGAN for unsupervised paired data synthesis before training the diffusion model. However, this reliance limits its performance to be similar to CycleGAN.  Li et al.~\cite{li2023zero} proposed a novel framework for zero-shot medical image translation based on the diffusion model, which only requires target domain images for training. However, it is built on the assumption that the source domain and the target domain share most of the low- and high-frequency information, which is not true in the MR-to-CT translation task. In general, GANs and VAEs can better preserve anatomical structures through unsupervised training, while diffusion models, though challenging to train in an unsupervised manner, generate more realistic and higher-quality images. The development of FDDM combines the advantages of VAEs in achieving anatomical structure conversion in an unsupervised manner with the strengths of diffusion models in generating high-quality images.

\begin{figure*}[ht]
\centering
  \includegraphics[width=.99\textwidth]{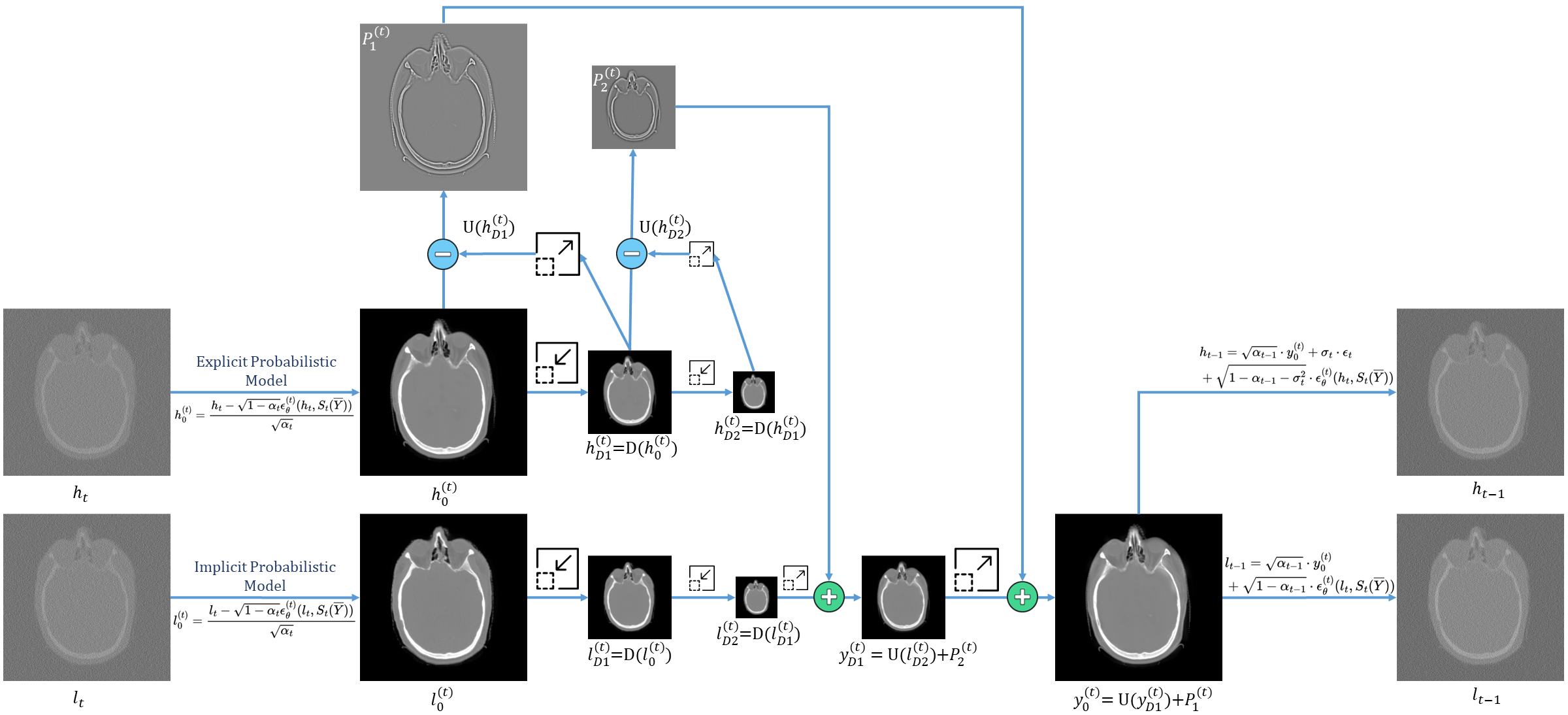}
  \caption{This figure shows the detailed framework of our dual-path reverse diffusion, where $h_t$ and $l_t$ are the noisy CT images at step $t$ for the explicit probabilistic model and the implicit probabilistic model, respectively. The clean CT images $h_0^{(t)}$ and $l_0^{(t)}$ are predicted by the models and fused using the Laplacian pyramid, resulting in $y_0^{(t)}$. Subsequently, $h_{t-1}$ and $l_{t-1}$ are obtained through different paths.}
  \label{fig:details_structure}
\end{figure*}

\begin{figure}[ht]
\centering
  \includegraphics[width=.5\textwidth]{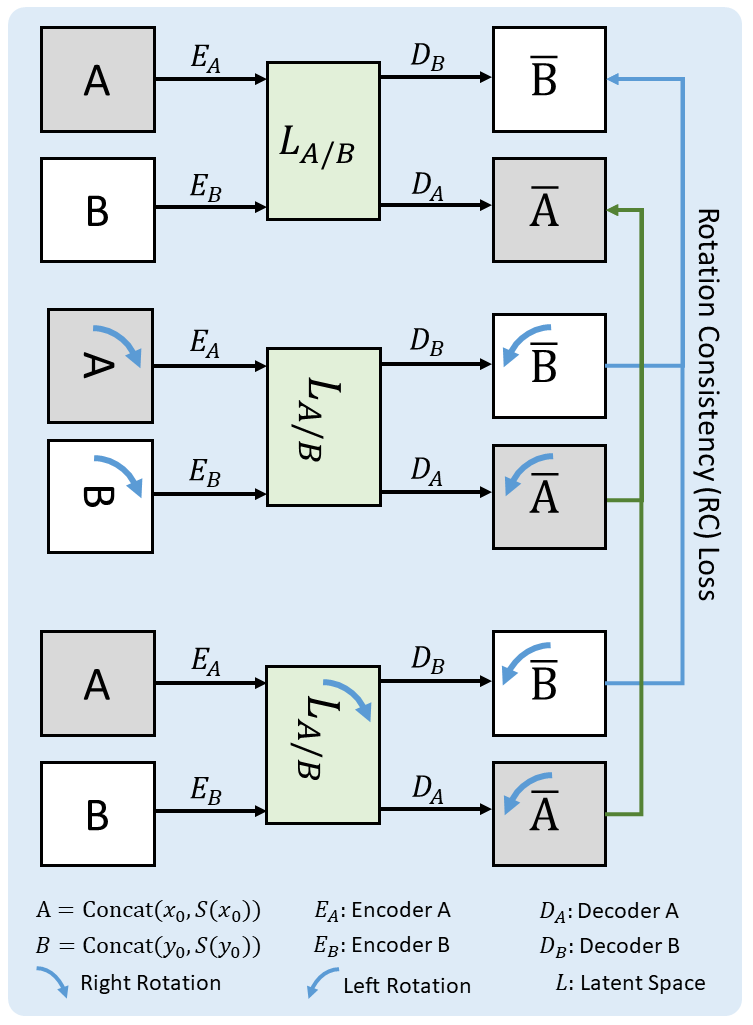}
  \caption{Diagram of the initial conversion module and the rotation consistency loss.}
  \label{fig:init}
\end{figure}

\section{Method}
The overall workflow of our method is shown in Fig. \ref{fig:brief}. Given a source MR image $X$, we employ an unsupervised initial conversion module based on VAE to obtain the CT anatomical information, including CT boundary (local structure) $S(\overline{Y})$ and noisy CT (overall pattern) $y_{T_s}$. Subsequently, using the anatomical information of the CT, the dual-path reverse diffusion process generates the high-quality CT image $y_0$.

\subsection{Initial Conversion Module}
Since diffusion models are challenging to train directly in an unsupervised manner, we first use VAE to convert the MR to the corresponding anatomical information of the CT, which is then used as a condition in the generation process of the diffusion model. First, we define the anatomical information of the CT to include both the local structure and the overall pattern. The overall pattern is obtained through the initial conversion module followed by forward diffusion. Below are the training details of the VAE-based initial conversion module.

First, we use the Sobel filter to extract the MR boundary $S(X)$ and CT boundary $S(Y)$ from unpaired MR image $X$ and CT image $Y$. Then, we define the inputs of the initial conversion module as A and B:

\begin{equation}
A = \text{Concat}(X, S(X)),
\end{equation}

\begin{equation}
B = \text{Concat}(Y, S(Y)).
\end{equation}

As shown in Fig.~\ref{fig:init}, our initial conversion module comprises two encoders $E_A$ and $E_B$ and two decoders $D_A$ and $D_B$, based on the assumption of a shared latent space $L_{A/B}$ between the two encoders and two decoders. This latent space can be independently derived from each domain and can be used to regenerate the MR or CT from the latent space, shown as:

\begin{align}
&L_{A/B} = L_A = E_A (A) = L_B = E_B(B), \\
&\overline{B} = D_B (L_{A/B}), \\
&\overline{A} = D_A (L_{A/B}),
\end{align}

where $\overline{A}$ and $\overline{B}$ are the synthesized MR and CT by the decoder. To better describe our VAE training process, we first define the variational posterior distributions $q_A (L_A|A)$ and $q_B (L_B|B)$ as:

\begin{align}
&q_A (L_A|A) = \mathcal{N}(L_A |E_{\mu,A}(A),I), \\
&q_B (L_B|B) = \mathcal{N}(L_B | E_{\mu,B}(B),I),
\end{align}

where $E_{\mu,A}(A)$ and $E_{\mu,B}(B)$ denote the mean output of the encoders, and $I$ is the identity matrix. A universal and essential component in VAE training is minimizing the variational upper bound. The loss function for this training is as follows:

\begin{equation}
\begin{split}
\mathcal{L}_{\text{\tiny VAE}_A}=&\lambda_1 \text{KL}( q_A(L_A|A) || \mathcal{N}(0,I) ) \\
&+ \lambda_2 \mathbb{E}_{L_A \sim q_A (L_A|A)}[|A-D_A(L_A)|],
\end{split}
\end{equation}

\begin{equation}
\begin{split}
\mathcal{L}_{\text{\tiny VAE}_B}=&\lambda_1 \text{KL}( q_B(L_B|B) || \mathcal{N}(0,I) ) \\
&+ \lambda_2 \mathbb{E}_{L_B \sim q_B (L_B|B)}[|B-D_B(L_B)|],
\end{split}
\end{equation}

where $\lambda_1$ and $\lambda_2$ denote hyperparameters, $\mathbb{E}$ represents the expectation, and $\mathcal{N}(0,I)$ is a Gaussian distribution with mean 0. In simple terms, the training objective is to make the latent $L_{A/B}$ conform to a Gaussian distribution and minimize the reconstruction error after encoding and decoding. However, the above training alone is not sufficient to achieve unsupervised image translation capabilities, which necessitates the incorporation of discriminator loss functions represented as:

\begin{equation}
\begin{split}
\mathcal{L}_{\text{\tiny Dis}_A}&=\mathbb{E}_{A \sim P_{\mathcal{A}}} [\log Dis_A (A) ]\\
&+\mathbb{E}_{L_B \sim q_B (L_B|B)}[\log(1-Dis_A(D_A(L_B)))],
\end{split}
\end{equation}

\begin{equation}
\begin{split}
\mathcal{L}_{\text{\tiny Dis}_B}&=\mathbb{E}_{B \sim P_{\mathcal{B}}} [\log Dis_B (B) ]\\
&+\mathbb{E}_{L_A \sim q_A (L_A|A)}[\log(1-Dis_B(D_B(L_A)))],
\end{split}
\end{equation}
where $Dis_A$ and $Dis_B$ represent the discriminators, and $P_{\mathcal{A}}$ and $P_{\mathcal{B}}$ are marginal data distributions for MR $A$ and CT $B$, respectively. In unsupervised image translation tasks, we do not have paired (MR, CT) images, and we can only utilize the marginal distributions of their respective datasets. These equations represent the most commonly used GAN losses, aimed at making the generated results conform more closely to the given data distribution through the discriminator. We also used the cycle consistency loss commonly used in unsupervised image translation:

\begin{equation}
\begin{split}
\mathcal{L}_{\text{\tiny CC}_A}&=\lambda_1\text{KL}( q_A(L_A|A) || \mathcal{N}(0,I) )\\
&+\lambda_1\text{KL}( q_B(L_B|D_B(E_A(A)))) || \mathcal{N}(0,I) )\\
&+\lambda_2\mathbb{E}_{L_B \sim q_B (L_B|D_B(E_A(A)))}[|A-D_A(L_B)|],
\end{split}
\end{equation}

\begin{equation}
\begin{split}
\mathcal{L}_{\text{\tiny CC}_B}&=\lambda_1\text{KL}( q_B(L_B|B) || \mathcal{N}(0,I) )\\
&+\lambda_1\text{KL}( q_A(L_A|D_A(E_B(B)))) || \mathcal{N}(0,I) )\\
&+\lambda_2\mathbb{E}_{L_A \sim q_A (L_A|D_A(E_B(B)))}[|B-D_B(L_A)|].
\end{split}
\end{equation}

Additionally, we designed a rotation consistency loss to further enhance the accuracy of the model outputs. The core idea is to ensure that the generated images remain consistent under different rotational transformations to the images or latent space, which helps the model better preserve the structural information of the images. By defining the rotation of an image or latent space 90 degrees to the left as $LR(\cdot)$ and 90 degrees to the right as $RR(\cdot)$, the rotation consistency losses are formulated as:

\begin{equation}
\begin{split}
\mathcal{L}_{\text{\tiny RC}_A}=& \lambda_2 \mathbb{E}[|LR(D_B(E_A(RR(A))))-D_B(E_A(A))|] \\
&+ \lambda_2 \mathbb{E}[|LR(D_B(RR(E_A(A))))-D_B(E_A(A))|],
\end{split}
\end{equation}

\begin{equation}
\begin{split}
\mathcal{L}_{\text{\tiny RC}_B}=& \lambda_2 \mathbb{E}[|LR(D_A(E_B(RR(B))))-D_A(E_B(B))|] \\
&+ \lambda_2 \mathbb{E}[|LR(D_A(RR(E_B(B))))-D_A(E_B(B))|].
\end{split}
\end{equation}

The final loss function of our initial conversion module is the sum of all the above losses. Through the trained initial conversion module, we can convert MR-domain inputs $X$ and $S(X)$ to obtain coarse CT $\overline{Y}$ and CT boundary $S(\overline{Y})$, as follows:

\begin{equation}
\begin{split}
\text{Concat}(\overline{Y}, S(\overline{Y})) = \overline{B} = D_B (E_A(A)).
\end{split}
\end{equation}

Subsequently, we perform a forward diffusion (i.e., low-pass filtering) to convert the coarse CT $\overline{Y}$ to the noisy CT $y_{T_s}$ as the starting point for reverse diffusion. As shown in Fig. \ref{fig:brief}, $S(\overline{Y})$ represents the local structure, and $y_{T_s}$ represents the overall CT anatomy pattern. Together, they serve as the necessary anatomical information of the CT to guide the reverse diffusion process. Even though we already have a coarse CT $\overline{Y}$, the following diffusion process is still necessary for further image correction and enhancement, since the image quality of the coarse CT $\overline{Y}$ from the VAE-based initial conversion module is low with data distribution mismatches. The related experimental validation is presented in Section \ref{ablation}.

\subsection{Forward Diffusion Process}
The following introduces a detailed background on diffusion models and elucidates how low-pass filtering is accomplished through forward diffusion. For the diffusion model~\cite{sohl2015deep,ho2020denoising,song2021denoising}, the forward process $q_D(y_t | \overline{Y})$ with Gaussian noise is commonly defined as:
\begin{equation} \label{q_d}
q_D(y_t | \overline{Y}) = \mathcal{N}(\sqrt{\alpha_t} \overline{Y}, (1 - \alpha_t) I),
\end{equation}
where $y_t$ denotes the noise-corrupted image at step $t$. The noise schedule sequence $\alpha_{1:t}$ decreases monotonically and stays within $(0, 1]^T$. Considering that Gaussian noise has the same intensity at all frequencies, in this study we use blue noise to achieve an effect closer to low-pass filtering through forward diffusion. Blue noise is characterized by linearly increasing intensity with frequency, equipping it with more energy in higher frequencies than in lower frequencies. To incorporate blue noise into the forward process, we modify the noise term in equation \eqref{q_d} as: 

\begin{equation} \label{e_func1}
q_D^{Blue}(y_t | \overline{Y}) = \mathcal{N}_{Blue}(\sqrt{\alpha_t} \overline{Y}, (1 - \alpha_t) I).
\end{equation}
Following this equation, we can represent $y_t$ as:

\begin{equation} \label{y_t_from_y}
y_t = \sqrt{\alpha_t} \overline{Y} + \sqrt{1 - \alpha_t} z_b, \quad  \quad z_b \sim \mathcal{N}_{blue}(0, I)
\end{equation}
with $z_b$ represents the blue noise sampled from a distribution $\mathcal{N}_{blue}(0, I)$, which has a power spectral density (PSD) linearly increasing with frequency:
\begin{equation}
\text{PSD}_{blue}(f) = k \cdot f,
\end{equation}
where $k$ is the proportional constant, and $f$ is the frequency. Correspondingly, the noise power is higher at higher frequencies, leading to more attenuation of high-frequency components. Consequently, the process mimics the effect of a low-pass filter, where lower frequency components are relatively preserved while higher frequency components are more reduced. Therefore, incorporating blue noise into the forward diffusion process effectively approximates a low-pass filter. Serving a low-pass filter, the forward diffusion process helps to disrupt the original image/information and yield new image/information through a reverse diffusion process to correct the errors in the original image/information.

Knowing that the forward diffusion process approximates a low-pass filter, a to-be-determined parameter is the threshold for the low-pass filter, i.e., how many steps the forward diffusion should take. Since $\alpha_t$ is a decreasing sequence, $\text{SNR}(y_t, f)$ decreases as the forward diffusion step $t$ increases. Therefore, if the number of steps is too large, very little image information would be retained to effectively guide the reverse diffusion process. On the other hand, if the number of steps is too small, it won't adequately disrupt/override the data distribution errors in the coarse CT $\overline{Y}$ to allow further image correction and enhancement. We hypothesize that there is an optimal number of steps $T_s$, which will be determined through experiments (as discussed in Section \ref{sec:T_s}).

\subsection{Dual-Path Reverse Diffusion Process}
We previously defined the VAE-converted CT anatomical information, which includes both local structures and overall patterns of the CT images. To better use the two types of anatomical information, they are integrated into the model guidance in different ways by FDDM. The CT boundary information, $S(\overline{Y})$, representing local structures of the CT images, is incorporated as additional conditional input to the network. In parallel,  $y_t$, which represents the overall CT anatomical pattern after the low-pass filtering, serves as the starting point for the reverse diffusion process. Therefore, the overall pattern of the CT image will continuously evolve throughout the reverse diffusion process, which is conditioned by $S(\overline{Y})$. Compared with high-frequency information including the anatomy boundary and local structures, the overall pattern and appearance of CT images translated by the VAE model, primarily consisting of low-frequency information, are relatively more reliable and should not be substantially altered by the diffusion model. Thus, we designed a dual-path reverse diffusion process for low and high frequencies, which is detailed below.

Reverse diffusion denotes the noise-predicting and gradual denoising process of the diffusion model to obtain a clean image. Specifically, it predicts $y_0^{(t)}$ from $y_t$ and CT boundary $S_t(\overline{Y})$ at each step $t$  to facilitate the transition from $y_t$ to $y_{t-1}$. As the number of steps decreases, the image noise reduces, and the need for CT boundary information diminishes. Correspondingly, the threshold gradually increases, progressively eliminating the guidance from CT boundary information. On the other hand, gradually phasing out the high-frequency condition from the VAE conversion module allows the diffusion model to further correct the residual high-frequency errors inherent to the VAE conversion. Therefore, at each step $t$, we apply the threshold $\frac{T_s - t}{T_s}$ to the [0, 1]-normalized CT boundary information $S(\overline{Y})$ (setting values below this threshold to zero) to obtain the corresponding new CT boundary $S_t(\overline{Y})$. For the reverse diffusion step $t$ of the noisy image $y_t$, we predict the noiseless $y_0^{(t)}$ at each step using the following equation:

\begin{equation}
y_0^{(t)} = \frac{y_t - \sqrt{1-\alpha_t} \epsilon_\theta^{(t)}(y_t,S_t(\overline{Y}))}{\sqrt{\alpha_t}},
\label{eq2}
\end{equation}

where $\epsilon_\theta^{(t)}(y_t, S_t(\overline{Y}))$ is the noise predicted by the diffusion model based on $y_t$ and $S_t(\overline{Y})$. The reverse diffusion process from step $t$ to $t-1$ is:

\begin{equation}
\begin{aligned}
&y_{t-1} = \sqrt{\alpha_{t-1}} \cdot \underbrace{\frac{y_t - \sqrt{1 - \alpha_t} \epsilon_\theta^{(t)}(y_t,S_t(\overline{Y}))}{\sqrt{\alpha_t}} }_{\text{predicted } y_0^{(t)} } \\& + \underbrace{ \sigma_t \cdot \epsilon_t + \sqrt{1 - \alpha_{t-1} - \sigma_t^2} \cdot \epsilon_\theta^{(t)}(y_t,S_t(\overline{Y}))}_{\text{noise of } y_{t-1}}.
\end{aligned}
\label{eq3}
\end{equation}

$\epsilon_t$ denotes a random noise introduced at each step. When the coefficient $\sigma_t$ for the random noise $\epsilon_t$ is 0, the random term is removed, and the process becomes deterministic. Correspondingly, the model becomes an implicit probabilistic model. Alternatively, when $\sigma_t = \sqrt{\frac{1 - \alpha_{t-1}}{1 - \alpha_t}} \sqrt{1 - \frac{\alpha_t}{\alpha_{t-1}}}$, the model is the same as DDPM, rendering an explicit probabilistic model. The explicit probabilistic model (DDPM) adds random noise to the image at each step, introducing stochasticity to the process to capture a broader potential output distribution~\cite{karras2024analyzing}. In the implicit probabilistic model, by eliminating the random noise, the model transitions between steps deterministically based only on the learned parameters and initial conditions. This deterministic approach can achieve a more stable reverse diffusion process. Since we do not expect excessive changes in the low-frequency information (overall anatomical pattern) that could adversely affect anatomical accuracy, we use the implicit probabilistic reverse diffusion model for low-frequency information. In contrast, to allow the diffusion model to better correct the erroneous anatomical boundary information and local structure errors, we use the explicit probabilistic model for high-frequency information to allow more changes in the reverse diffusion process. This dual-path design helps maintain the anatomical fidelity of the CT images while allowing the necessary variability to accurately model the data distribution.

Considering that both explicit and implicit probabilistic models operate on the complete image rather than directly on a specific frequency band, we designed a method where the results of each step are predicted through both models and then fused and reconstructed using the Laplacian pyramid~\cite{wang2020multi} to achieve predictions for low and high frequencies through different models. In our equations, $y_{t}$ represents the noisy CT at step $t$. To further distinguish between the two paths, we use $h_{t}$ and $l_{t}$ to represent the explicit probabilistic model for high-frequency information and the implicit probabilistic model for low-frequency information, respectively. Both paths take $y_{T_s}$ as the starting point, i.e., $h_{T_s} = l_{T_s} = y_{T_s}$. Then, at any reverse diffusion step $t$, we predict the clean CT images $h_0^{(t)}$ and $l_0^{(t)}$ using the following equations:

\begin{equation}
h_0^{(t)} = \frac{h_t - \sqrt{1-\alpha_t} \epsilon_\theta^{(t)}(h_t,S_t(\overline{Y}))}{\sqrt{\alpha_t}},
\label{eq2}
\end{equation}

\begin{equation}
l_0^{(t)} = \frac{l_t - \sqrt{1-\alpha_t} \epsilon_\theta^{(t)}(l_t,S_t(\overline{Y}))}{\sqrt{\alpha_t}}.
\label{eq2}
\end{equation}

After obtaining $h_0^{(t)}$ and $l_0^{(t)}$, we calculate their Gaussian pyramids to combine the low-frequency information of $l_0^{(t)}$ and the high-frequency information of $h_0^{(t)}$ . For the Laplacian pyramid, the Gaussian blur is applied using a Gaussian kernel, and the downsample operation reduces the width and height of the image by half at each step:

\begin{equation}
h_{D1}^{(t)} = D(h_{0}^{(t)}) = \text{Downsample}(\text{GaussianBlur}(h_0^{(t)})),
\end{equation}
\begin{equation}
h_{D2}^{(t)} = D(h_{D1}^{(t)}) = \text{Downsample}(\text{GaussianBlur}(h_{D1}^{(t)})),
\end{equation}
\begin{equation}
l_{D1}^{(t)} = D(l_{0}^{(t)}) = \text{Downsample}(\text{GaussianBlur}(l_0^{(t)})),
\end{equation}
\begin{equation}
l_{D2}^{(t)} = D(l_{D1}^{(t)}) = \text{Downsample}(\text{GaussianBlur}(l_{D1}^{(t)})).
\end{equation}

From the resulting images, the Laplacian pyramid of $h_{0}^{(t)}$ is obtained by subtracting the upsampled Gaussian levels from the same-level images acquired through the down-sampling path, which extracts the high-frequency information $P_{1}^{(t)}$ and $P_{2}^{(t)}$. 

\begin{equation}
P_{1}^{(t)} = h_{0}^{(t)} - U(h_{D1}^{(t)}) = h_{0}^{(t)} - \text{Upsample}(h_{D1}^{(t)}),
\end{equation}
\begin{equation}
P_{2}^{(t)} = h_{D1}^{(t)} - U(h_{D2}^{(t)}) = h_{D1}^{(t)} - \text{Upsample}(h_{D2}^{(t)}).
\end{equation}

In above equations, the upsample operation increases the width and height of the image by a factor of two at each step. Then, we fuse the low-frequency information from $l_0^{(t)}$ and the high-frequency information from $h_0^{(t)}$ to reconstruct $y_0^{(t)}$ as follows:

\begin{equation}
y_{D1}^{(t)} = U(l_{D2}^{(t)})  + P_{2}^{(t)}  = \text{Upsample}(l_{D2}^{(t)})  + P_{2}^{(t)} 
\end{equation}
\begin{equation}
y_{0}^{(t)} = U(y_{D1}^{(t)})  + P_{1}^{(t)} = \text{Upsample}(y_{D1}^{(t)})  + P_{1}^{(t)} 
\end{equation}

Through the fused $y_0^{(t)}$, the low-frequency information predicted by the implicit probabilistic model and the high-frequency information predicted by the explicit probabilistic model are preserved. Next, the explicit probabilistic model converts $h_{t}$ to $h_{t-1}$ using the following equation:

\begin{equation}
\begin{aligned}
& h_{t-1} = \sqrt{\alpha_{t-1}} \cdot y_0^{(t)} + \sigma_t \cdot \epsilon_t  \\ & + \sqrt{1 - \alpha_{t-1} - \sigma_t^2} \cdot \epsilon_\theta^{(t)}(h_t,S_t(\overline{Y})) .
\end{aligned}
\label{eq3}
\end{equation}

The implicit probabilistic model converts $l_{t}$ to $l_{t-1}$ using the following equation:

\begin{equation}
\begin{aligned}
& l_{t-1} = \sqrt{\alpha_{t-1}} \cdot y_0^{(t)}   \\ & + \sqrt{1 - \alpha_{t-1} } \cdot \epsilon_\theta^{(t)}(l_t,S_t(\overline{Y})).
\end{aligned}
\label{eq3}
\end{equation}

Starting from the new step $t-1$, the Laplacian pyramid-based fusion steps as step $t$ were repeated to predict the images of step $t-2$, which forms an iterative loop until the final CT $y_0$ is obtained.

\section{Experiments}

\begin{table}[htb]
    \centering
    \caption{Comparison of different methods on the brain dataset. Downward arrows indicate "lower is better" and upward arrows indicate "higher is better". The results for FDDM are obtained based on $T_s = 300$. Values marked with * indicate statistical significance ($p < 0.05$) using the Wilcoxon signed-rank test compared to FDDM.}
    \begin{tabular}{lcccc}
        \toprule
        Method & FID $\downarrow$  & SSIM $\uparrow$ & MSE (HU) $\downarrow$ & MAE (HU) $\downarrow$ \\ 
        \toprule
        CycleGAN & 62.4051 & 0.8704* $\pm$ 0.0510 & 17999.873* $\pm$ 7474.353 & 39.261* $\pm$ 13.511 \\ 
        GcGAN & 60.0342 & 0.8786* $\pm$ 0.0504 & 19875.950* $\pm$ 10129.532 & 39.842* $\pm$ 16.025 \\ 
        RegGAN & 73.7575 & 0.8033* $\pm$ 0.0585 & 97734.234* $\pm$ 86313.220 & 102.163* $\pm$ 59.181 \\ 
        UNIT & 49.7074 & 0.8912* $\pm$ 0.0459 & 17454.355* $\pm$ 8741.541 & 37.034* $\pm$ 14.508 \\ 
        MUNIT & 72.8495 & 0.8338* $\pm$ 0.0646 & 30446.590* $\pm$ 14410.634 & 54.542* $\pm$ 20.573 \\ 
        SDEdit & 75.9913 & 0.7057* $\pm$ 0.1085 & 106480.305* $\pm$ 59253.810 & 118.759* $\pm$ 52.682 \\ 
        SynDiff & 70.5355 & 0.8503* $\pm$ 0.0553 & 42668.074* $\pm$ 21847.432 & 59.746* $\pm$ 23.925 \\ 
        \textbf{FDDM} & \textbf{25.8565} & \textbf{0.9089} $\pm$ \textbf{0.0417} & \textbf{12345.224} $\pm$ \textbf{7052.862} & \textbf{28.510} $\pm$ \textbf{12.651} \\ 
        \bottomrule
    \end{tabular}
    \label{tab:brain_results}
\end{table}

\begin{figure*}[htb]
\centering
  \includegraphics[width=.99\textwidth]{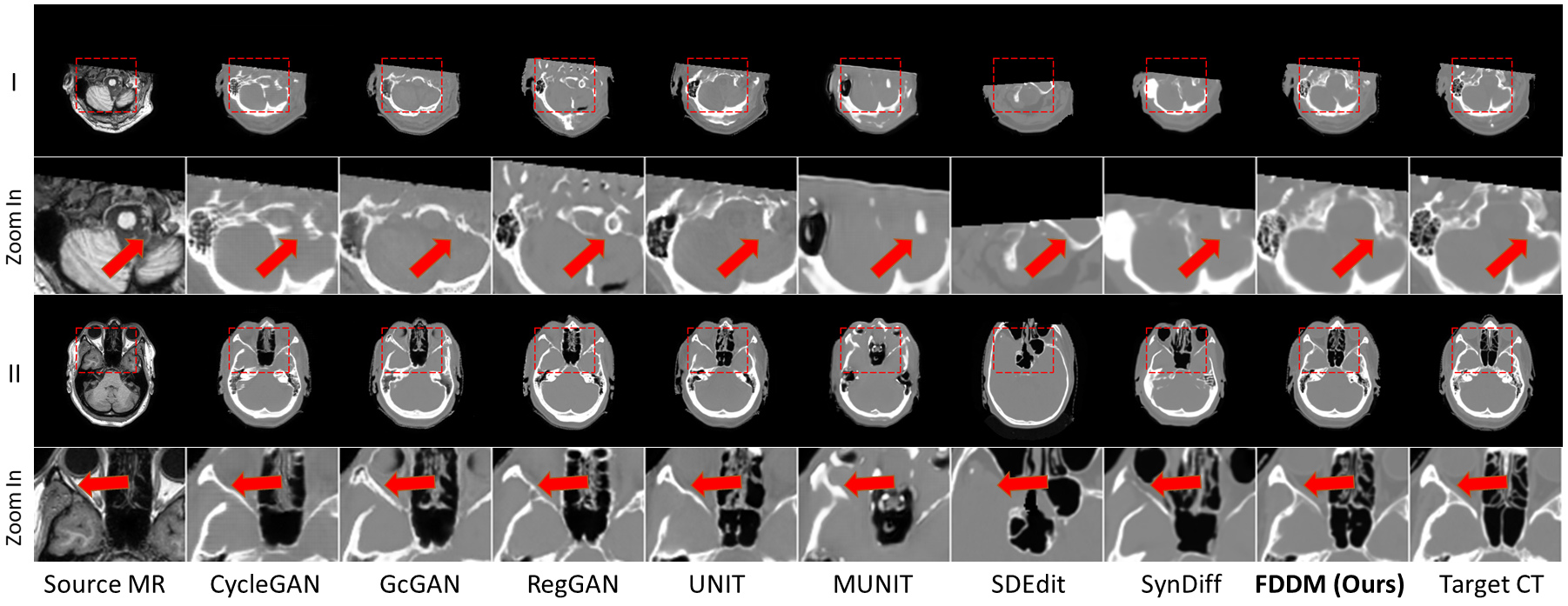}
  \caption{Visual comparison between other models and FDDM on the brain MR-to-CT translation dataset.}
  \label{fig:brain}
\end{figure*}

\subsection{Experimental and Dataset Setup}
In our study, we used the SynthRAD2023 dataset~\cite{thummerer2023synthrad2023,huijben2024generating}, which includes 180 pairs of 3D scans from the brain (MR and CT) and 180 pairs of 3D scans from the pelvis (MR and CT). For our experimental setup, we randomly split these images, reserving 10 pairs of CT and MR scans each from the brain and pelvis as our test set, and 10 pairs each for validation (per needs of comparison methods). The remaining 160 paired MR-CT scans were used for training. Each 3D image was processed into 2D slices. To simulate unpaired (unsupervised) training, we shuffled the slices and removed the pairing information from the training dataset. All slices were uniformly resized to a resolution of 256 × 256 pixels. For MR images, we applied intensity clipping, removing the bottom 0.5\% of intensity values. For CT images, the Hounsfield Unit (HU) values were clipped to the range of [-1000, 1000]. All images were then rescaled to [-1, 1]. The training MR and CT slices were used to train the initial conversion module, while only the training CT slices were used to train the diffusion model, which does not need any MR input. 
In our implementation, several parameters were determined based on default values from previous methods. Our diffusion model has $T=1000$ steps. The learning rates for the VAE-based initial conversion module and the diffusion model were set to $1 \times 10^{-4}$ and $2 \times 10^{-4}$, respectively. Exponential Moving Average (EMA) was applied to the model parameters with a decay factor of 0.9999. 
The loss weights for the initial conversion module followed the widely used VAE framework, with $\lambda_1$ set to 0.01 and $\lambda_2$ set to 10. The value of $T_{s}$ during the inference process was set to 500, based on parameter studies (Section III). We used the Pytorch library to implement the FDDM framework. Experiments were conducted on an NVIDIA A100 80G GPU card. We compared FDDM with other state-of-the-art methods (GcGAN, RegGAN, CycleGAN, UNIT, MUNIT, SynDiff, SDEdit) using official open-source code \cite{zhu2017unpaired,fu2019geometry,liu2017unsupervised,huang2018multimodal,meng2021sdedit}.

\begin{table}[htb]
    \centering
    \caption{Comparison of different methods on the pelvis MR-to-CT translation dataset. Downward arrows indicate "lower is better" and upward arrows indicate "higher is better". The results for FDDM are obtained based on $T_s = 300$. Values marked with * indicate statistical significance ($p < 0.05$) using the Wilcoxon signed-rank test compared to FDDM.}
    \begin{tabular}{lcccc}
        \toprule
        Method & FID $\downarrow$  & SSIM $\uparrow$ & MSE (HU) $\downarrow$ & MAE (HU) $\downarrow$ \\ 
        \toprule
        CycleGAN & 77.6318 & 0.8840* $\pm$ 0.0256 & 13001.760* $\pm$ 11438.179 & 30.140* $\pm$ 11.539 \\ 
        GcGAN & 43.7356 & 0.9014* $\pm$ 0.0257 & 12104.721* $\pm$ 11375.784 & 26.564* $\pm$ 11.674 \\ 
        RegGAN & 56.0113 & 0.8763* $\pm$ 0.0412 & 17782.205* $\pm$ 19752.328 & 34.128* $\pm$ 21.893 \\ 
        UNIT & 44.5953 & 0.8973* $\pm$ 0.0238 & 11522.925* $\pm$ 10758.339 & 26.997* $\pm$ 10.821 \\ 
        MUNIT & 74.3732 & 0.8726* $\pm$ 0.0248 & 15659.264* $\pm$ 11134.156 & 34.739* $\pm$ 11.051 \\ 
        SDEdit & 73.1082 & 0.6726* $\pm$ 0.1569 & 72268.590* $\pm$ 54065.650 & 115.527* $\pm$ 71.631 \\ 
        SynDiff & 74.6271 & 0.8687* $\pm$ 0.0255 & 16065.846* $\pm$ 10392.632 & 33.364* $\pm$ 10.664 \\ 
        \textbf{FDDM} & \textbf{29.2003} & \textbf{0.9117} $\pm$ \textbf{0.0235} & \textbf{8561.108} $\pm$ \textbf{10817.484} & \textbf{22.254} $\pm$ \textbf{10.944} \\ 
        \bottomrule
    \end{tabular}
    \label{tab:pelvis_results}
\end{table}

\begin{figure*}[htb]
\centering
  \includegraphics[width=.99\textwidth]{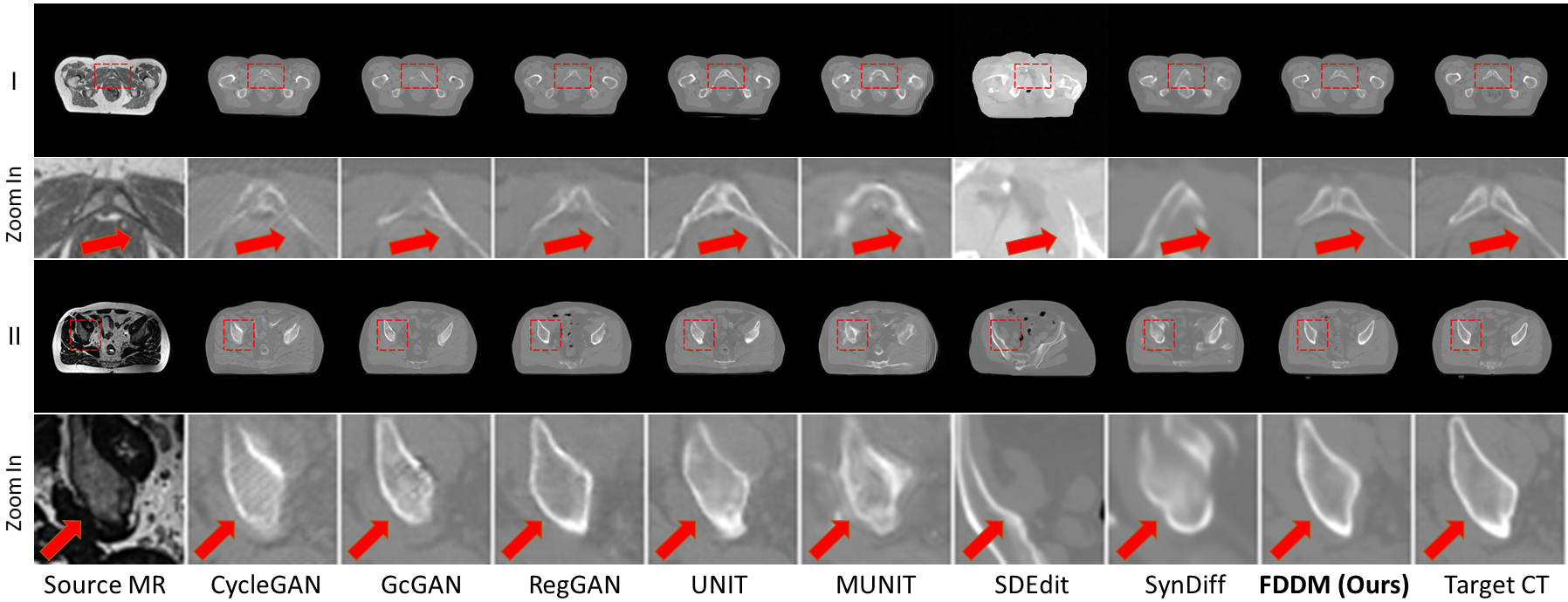}
  \caption{Visual comparison between other models and FDDM on the pelvis MR-to-CT translation dataset.}
  \label{fig:pelvis}
\end{figure*}

\subsection{Comparison with Other Methods}

To evaluate the performance of FDDM, we compared it with GAN-based (CycleGAN, GcGAN, and RegGAN), VAE-based (UNIT and MUNIT), and diffusion model-based (SDEdit and SynDiff) MR-to-CT conversion methods on the brain and pelvis datasets.

\subsubsection{Brain Dataset}

Table \ref{tab:brain_results} provides a quantitative comparison of FDDM with other image conversion methods on the brain dataset using FID, SSIM, MSE in CT Hounsfield unit (HU), and MAE in CT Hounsfield unit (HU). FDDM achieved an FID score of 25.8565, substantially outperforming other methods, demonstrating its capability to generate highly realistic images. SSIM, MSE, and MAE, which can be considered as measures of anatomical accuracy, also showed that FDDM achieved the best performance, consistent with the visual comparison results in Fig. \ref{fig:brain}. Both the quantitative and visual results indicated that SDEdit performed the worst. SDEdit uses a diffusion model trained on CT images and directly performs forward and reverse diffusion on the source MR image. Essentially, SDEdit uses the low-frequency information extracted from the source MR image (as the forward diffusion process approximates a low-pass filter) to infer the CT image, although its inference capability is gained from training with low-frequency CT image information. However, the low-frequency information between CT and MR images differs significantly, resulting in substantially degraded results from SDEdit. Another diffusion model-based method, SynDiff, showed moderate performance because it uses CycleGAN to synthesize paired MR-CT data to avoid the challenges of unsupervised training, gaining a relative advantage in structural preservation. However, introducing CycleGAN for paired CT synthesis imposed an upper limit on the imaging realism, which aligns with our experimental results showing that SynDiff's performance was close to but slightly worse than CycleGAN. The VAE-based UNIT achieved a performance second only to our model, supporting the rationale of using VAE as the backbone structure for our initial conversion module and applying the diffusion model to further correct the residual errors.

\begin{table}[ht]
    \centering
    \caption{Ablation study of the initial conversion module, where RC stands for rotation consistency. The results are for the coarse CT generated from the VAE-based initial conversion module in the first stage and have not undergone the second stage diffusion model.}
    \begin{tabular}{lcccc}
        \toprule
Step & FID $\downarrow$ & SSIM $\uparrow$ & MSE (HU) $\downarrow$ & MAE (HU) $\downarrow$ \\
\toprule
without RC & 48.6921 & 0.8980 $\pm$ 0.0454 & 14615.918 $\pm$ 7824.331 & 33.0155 $\pm$ 13.7888 \\
\textbf{with RC} & \textbf{39.8187}  & \textbf{0.9126} $\pm$ \textbf{0.0405} & \textbf{11755.669} $\pm$ \textbf{6842.154} & \textbf{28.2204} $\pm$ \textbf{12.5264} \\
\bottomrule
    \end{tabular}
    \label{tab:rc}
\end{table}

\begin{figure}[ht]
\centering
  \includegraphics[width=.5\textwidth]{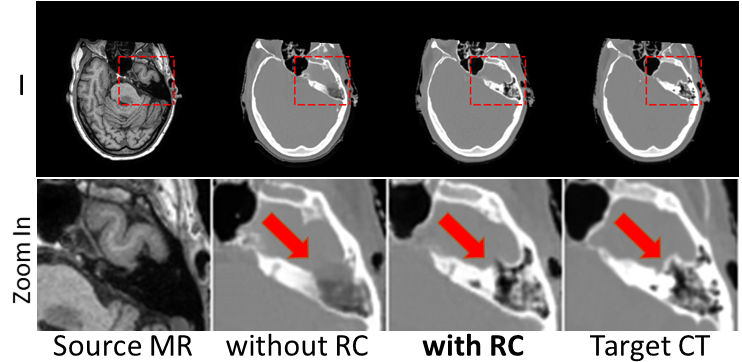}
  \caption{Visual results of the ablation study for the initial conversion module, where RC stands for rotation consistency. The images are from the coarse CTs obtained with the initial conversion module.}
  \label{fig:rc}
\end{figure}

\subsubsection{Pelvis Dataset}

In addition to the brain dataset, we also tested different methods on the pelvis dataset. The results in Table \ref{tab:pelvis_results} show that FDDM similarly achieved the best performance across all metrics. Overall, the performance of each model on the pelvis dataset was similar to that on the brain dataset, indicating stable performance. SDEdit continued to exhibit performance issues due to MR-CT domain gaps. For example, in sample I of Fig. \ref{fig:pelvis}, SDEdit generated CT images with incorrect intensity information. We also noticed that while both UNIT and MUNIT are VAE-based methods, UNIT outperformed MUNIT. It can be attributed to the fact that UNIT focuses on single-mode-based image conversion between two different domains. In contrast, MUNIT emphasizes multi-mode-based image conversion by decomposing the image into content and style codes, using independent content and style codes for diverse image generation. Although this design achieves multi-modal conversion beneficial for applications requiring diverse outputs, its performance in one-to-one mode conversion does not match that of UNIT. Specifically, the bony structure accuracy of MUNIT in Fig. \ref{fig:pelvis} is worse than the other methods.FDDM showed the highest accuracy in bony structures in Fig. \ref{fig:pelvis}, demonstrating its ability to correctly convert anatomical structures during image conversion. A comparison in the zoomed-in region revealed that FDDM achieved higher precision in very detailed bony structures than other methods.

\begin{table}[ht]
    \centering
    \caption{Ablation study of the second-stage diffusion model. The coarse CT results represent the outcomes of the VAE-based initial conversion module, before applying the diffusion model. IPM stands for the implicit probabilistic model, and EPM stands for the explicit probabilistic model, both used in our dual-path reverse diffusion process. These results are obtained based on $T_s = 300$. Since the results without CT boundary guidance are shown in Table \ref{tab:Ts}, we have not included them in the current table.}
    \begin{tabular}{lcccc}
        \toprule
        Step $\tilde{T}$ & FID $\downarrow$ & SSIM $\uparrow$ & MSE (HU) $\downarrow$ & MAE (HU) $\downarrow$ \\
        \toprule
        coarse CT & 39.8187 & 0.8916 $\pm$ 0.0534 & 14994.927 $\pm$ 8366.118 & 35.2006 $\pm$ 14.8481 \\
        without IPM & 26.6595 & 0.8839 $\pm$ 0.0558 & 14553.736 $\pm$ 7875.460 & 32.8674 $\pm$ 14.4395 \\
        without EPM & 27.9621 & 0.9020 $\pm$ 0.0462 & 13881.806 $\pm$ 7730.327 & 31.5993 $\pm$ 13.2413 \\
        \textbf{FDDM} & \textbf{25.8565} & \textbf{0.9116} $\pm$ \textbf{0.0409} & \textbf{11862.303} $\pm$ \textbf{6785.043} & \textbf{27.8005} $\pm$ \textbf{12.3200} \\
        \bottomrule
    \end{tabular}
    \label{tab:ipm_emp}
\end{table}

\begin{figure}[ht]
\centering
  \includegraphics[width=.5\textwidth]{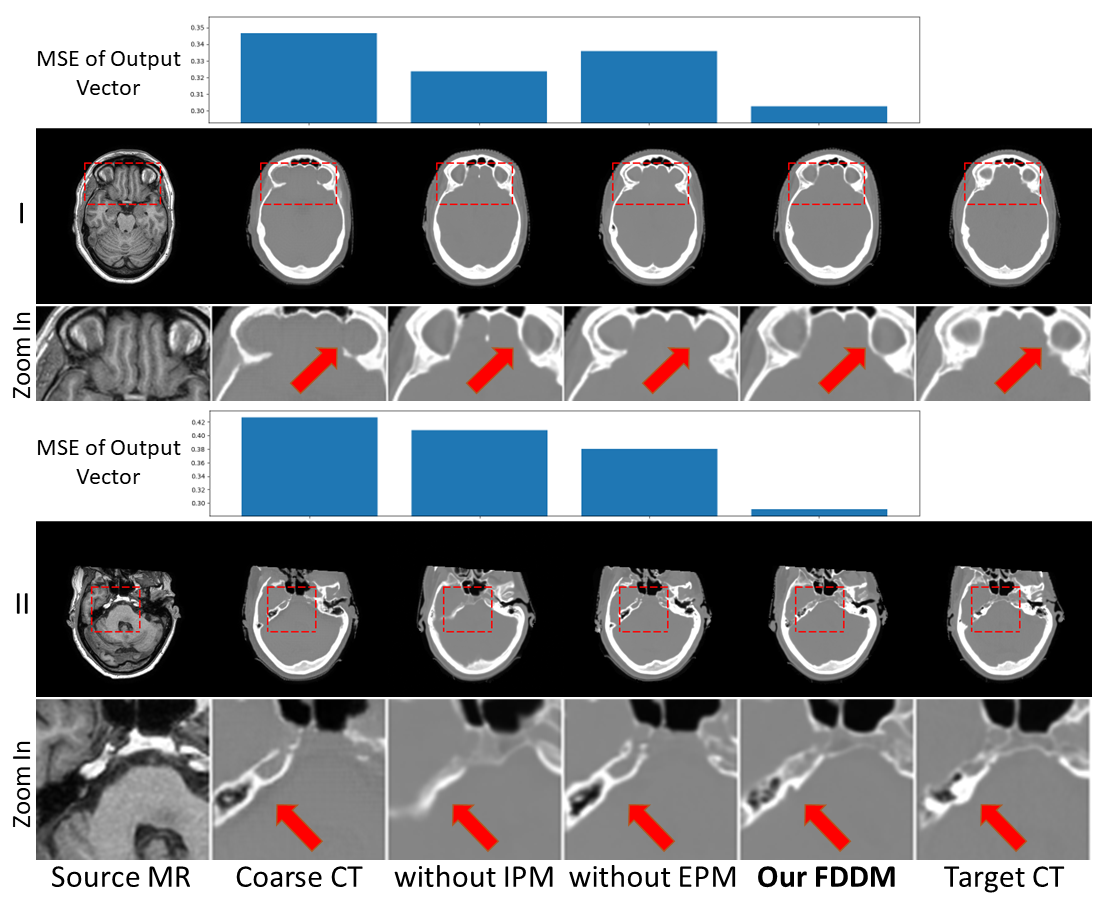}
  \caption{Visualization results of the ablation study for the second-stage diffusion model of FDDM.}
  \label{fig:ipm_epm}
\end{figure}

\subsection{Ablation Study}
\label{ablation}
To evaluate the efficacy of different components introduced into the two-staged FDDM framework, we first conducted ablation experiments to study the role of the rotation consistency loss used in the VAE-based initial conversion module. The results are shown in Table \ref{tab:rc}. When we removed rotation consistency, all metrics deteriorated. Fig. \ref{fig:rc} further corroborated the results by showing the converted CT has better anatomical accuracy with the rotation consistency loss. The results support our hypothesis that by enforcing the consistency of generated results under different transformations to the images or latent space, we can stabilize the model's output and reduce the hallucinations of anatomical structures. Since the second stage of FDDM relies on information derived from the initial conversion module to generate the final CT outputs, the accuracy of the initial conversion module is crucial.

In the second stage of FDDM, the diffusion model utilizes its learned prior knowledge of CT images to correct discrepancies in the data distributions of the initial results from the VAE-based first stage. The quantitative results of the ablation study for the second-stage diffusion model are shown in Table \ref{tab:ipm_emp}, and the visual results are in Fig. \ref{fig:ipm_epm}. Occasionally, differences in data distributions may not be easily noticeable to the naked eye, but a trained model can be very sensitive to them. For example, a trained classification or segmentation network's performance can degrade with poor data distribution. The FID calculation method is based on the distribution differences between generated and real images, and a higher FID value reflects a mismatch in data distribution. After optimization with the diffusion model, the FID value of the generated images decreases, indicating improved image quality and data distribution. To further validate our method, we compared the vector space obtained from a pre-trained Inception network with the ground truth CT's mean squared error (MSE) in Fig. \ref{fig:ipm_epm}. It can be seen that the results obtained after the diffusion model show a significant improvement in data distribution compared to the coarse CT.

\begin{table}[ht]
    \centering
    \caption{FDDM performance by using different diffusion steps $T_s$ during testing. Since our forward diffusion process approximates a low-pass filter, $T_s$ can be regarded as the low-pass filter threshold and its impact can be quantified.}
    \begin{tabular}{lcccc}
        \toprule
        Step $\tilde{T}$ & FID $\downarrow$ & SSIM $\uparrow$ & MSE (HU) $\downarrow$ & MAE (HU) $\downarrow$ \\
        \toprule
        100 & 28.2487 & \textbf{0.9136} $\pm$ \textbf{0.0401} & \textbf{11748.454} $\pm$ \textbf{6820.354} & \textbf{27.6307} $\pm$ \textbf{12.3094} \\
        \textbf{300} & \textbf{25.5146} & 0.9089 $\pm$ 0.0417 & 12345.224 $\pm$ 7052.862 & 28.5101 $\pm$ 12.6511 \\
        600 & 27.1035 & 0.8808 $\pm$ 0.0527 & 17777.790 $\pm$ 9870.868 & 36.0422 $\pm$ 16.1415 \\
        900 & 37.3593 & 0.7667 $\pm$ 0.0965 & 54486.934 $\pm$ 25457.920 & 78.4379 $\pm$ 31.5931 \\
        \bottomrule
    \end{tabular}
    \label{tab:Ts}
\end{table}

\begin{figure*}[ht]
\centering
  \includegraphics[width=.9\textwidth]{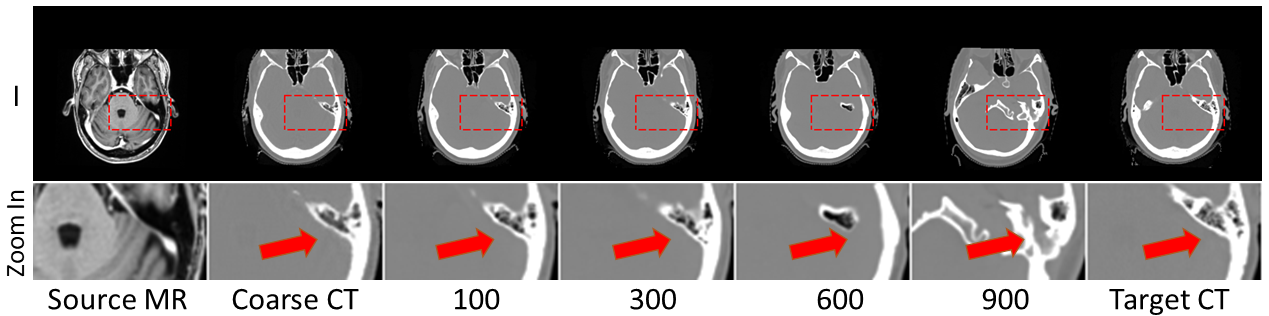}
  \caption{Visual results by using different diffusion steps $T_s$ during testing.}
  \label{fig:Ts}
\end{figure*}

Additionally, our diffusion model employs a dual-path reverse diffusion process, using an implicit probabilistic model without random noise for low frequencies and an explicit probabilistic model with random noise for high frequencies. The relevant results are shown in Table \ref{tab:ipm_emp}. When the implicit probabilistic model is removed, leaving only the explicit probabilistic model with random noise, SSIM, MSE, and MAE decrease while FID is more stable, indicating a decline in anatomical accuracy. It suggests that removing the implicit probabilistic model primarily affects anatomical accuracy as using only the explicit probabilistic model introduces additional randomness. However, when the explicit probabilistic model is removed, FID increases more, indicating a deterioration in image quality. During the reverse diffusion process, the model transitions from step $t$ to $t-1$ by predicting noise to obtain a clean CT and then adding the predicted noise with a smaller coefficient. This forms the basis of the diffusion model, allowing each diffusion step to remain stable. However, typically in explicit probabilistic models (such as DDPM), random Gaussian noise is also added at each step as a perturbation term to disrupt the existing image. By perturbing the existing information, the model can generate new information closer to the CT. Thus when the explicit probabilistic model is removed, the perturbation power of the diffusion model is reduced which degrades the FID. The same observation can be made in Fig. \ref{fig:ipm_epm}. In the first example of Fig. \ref{fig:ipm_epm}, the coarse CT has erroneous anatomical structures around the eyes. The results without the explicit probabilistic model are very close to the coarse CT, indicating that the model did not correct the errors. In contrast, the results with the explicit probabilistic model corrected the errors in the coarse CT, making the anatomical structures closer to the ground truth. However, the second example in Fig. \ref{fig:ipm_epm} shows that using only the explicit probabilistic model can introduce excessive randomness, leading to erroneous generation. Therefore, our dual-path reverse diffusion aims to strike a balance, correcting the errors in the coarse CT while preventing the model from generating overly randomized results.

\subsection{Impact of the Forward Diffusion Step $T_s$}
\label{sec:T_s}
For FDDM, although the diffusion model is trained with 1000 forward and reverse diffusion steps, we can customize this number of steps during testing which allows the diffusion model to act upon varying degrees of low-frequency information. The forward diffusion process approximates a low-pass filter, and the more steps, the less image information is retained. If the number of forward diffusion steps is too small, errors from the coarse CT image obtained by VAE conversion will be overly retained , affecting the final CT accuracy. However, if the number of steps is too large, too little image information is retained to effectively guide the reverse diffusion process. Therefore, we need to find an optimal $T_s$. In Table \ref{tab:Ts} and Fig. \ref{fig:Ts}, we tested the model's performance with different $T_s$. After comparison, we chose $T_s=300$ for our experimental setting. To minimize the compounding factors, we did not include CT boundary guidance information when testing the performance of different $T_s$. By comparing the results of $T_s=300$ in Table \ref{tab:Ts} and Table \ref{tab:ipm_emp}, we found that the FID score is slightly better without CT boundary guidance, but at the cost of a significant decrease in SSIM, MSE, and MAE when $T_s$ is large. The lower FID with CT boundary information-based guidance is because the CT boundary information is generated by the initial conversion module, which contains imperfections slightly affecting the image quality of final output CT. However, the high-frequency CT boundary information serves as strong guidance to preserve the anatomical accuracy as demonstrated in the previous work~\cite{li2023zero}. As shown in Table \ref{tab:Ts}, without the CT boundary guidance the structure integrity deteriorates quickly when the high-frequency information is removed with relatively large $T_s$ values.

\section{Conclusion and Future Work}

In this study, we introduced the Frequency Decoupled Diffusion Model (FDDM), a novel method for unpaired MR-to-CT image conversion in medical imaging. By combining the structural preservation advantages of VAE-based models and the high-fidelity image generation capabilities of diffusion models, along with FDDM's unique design of anatomical structure guidance and dual-path reverse diffusion, the model achieved high anatomical structure accuracy and high-quality image distribution for MR-to-CT conversion. A comprehensive evaluation of FDDM demonstrated its superior performance compared to state-of-the-art GAN-, VAE-, and diffusion-based models, paving the way for more precise disease diagnosis, localization, and treatment planning.

However, our current work has some limitations that need to be addressed in future work. Firstly, it is worth investigating the improvement in clinical applications of MR-generated CTs, such as the impact on the accuracy of treatment planning and dose calculations. Additional downstream clinical tasks, such as segmentation and diagnosis, can also be evaluated using the converted CTs. Secondly, the current FDDM model is implemented in 2D slice-by-slice due to efficiency considerations and memory limitations, while 3D diffusion models may better capture the volumetric dependency of anatomies to yield more accurate results. Lastly, our current model can only achieve one-to-one modality conversion. In future work, developing a universal model capable of conversions between multiple modalities can introduce more flexibility, robustness, and utility.

\bibliographystyle{ieeetr}
\bibliography{myrefs}

\end{document}